\documentclass[floats,preprint,prb,aps]{revtex4}
\begin{document}
\title{Generalized Coherent State Derivation of TDDFT Equations for Superconductors}
\author{Oleg Berman$^{1}$ and Shaul Mukamel$^{1,2}$}
\affiliation{Department of Chemistry$^{1}$ and Physics and Astronomy$^{2}$ \\
University of Rochester, Box 270216, Rochester, New York
14627-0216}

\vspace{0.5cm} \preprint{{\em Submitted to Phys. Rev. B.}
\hspace{3.7in} Web galley}
\date{\today}
\vspace{2.7in}

\begin{abstract}
Equations of motion are derived for the normal and the anomalous
single-electron density matrices of a Fermi liquid using
 a time dependent finite temperature generalized coherent state
(GCS) variational ansatz for the many-body density matrix.
Self-consistent equations for the order parameter $\Delta$ allow
to investigate the interplay of Coulomb repulsion and pairing
attraction in homogeneous and inhomogeneous Fermi-liquids with
spontaneously broken symmetry  such as
 high temperature superconductors. The
temperature of the Kosterlitz-Thouless transition to the
two-dimensional superfluidity is calculated.

\vspace{1.0 in}



\end{abstract}
\maketitle

\vspace{6mm}

\section{Introduction}

The BCS ansatz for the ground state wavefunction of
superconductors predicts a gap in the spectrum \cite{Schrieffer}
originating from the pairing of two electrons with opposite
momenta and spin projections. This pairing is induced by the
effective attraction induced by electron-phonon interaction. The
standard BCS description of superconductivity, as well as
Eliashberg's extension  to incorporate Coulomb repulsion between
electrons\cite{Eliashberg,mahan} do not apply for
strongly-correlated electrons such as in high-temperature
superconductors ($YBCO$)\cite{Carlson}. In two-dimensional ($2D$)
superconductors (e.g., in cuprates)\cite{Carlson,Varlamov}, the
phase transition to the superfluid state takes place at
temperatures below the mean-field phase transition temperature for
the appearance of the gap\cite{Kosterlitz,Nelson}, similar to the
Kosterlitz-Thouless phase transition. The important role played by
electron-electron correlations in the local order parameter above
the phase transition temperature, while the system is still in the
normal phase, was discussed\cite{Varlamov}.

Superconductivity in strongly correlated electron systems is often
described using effective Hubbard Hamiltonians\cite{Carlson}.
Taking electron-electron exchange and correlations into account,
the ground state energies and the collective excitation spectrum
of superconductors have been calculated by Density Functional
Theory (DFT) or its Time-Dependent extension
(TDDFT)\cite{Oliveira,Kohn,Grossbook,Gross}. In the
Oliveira-Gross-Kohn (OGK)  DFT
equations\cite{Oliveira,Kohn,Grossbook,Gross} the normal $\rho
(\mathbf{r},\mathbf{r}')$ and the anomalous $\sigma
(\mathbf{r},\mathbf{r}')$ density matrices
 satisfy the generalized
Bogoliubov-de Gennes equations\cite{Grossbook,deGennes} which
include one external field $v_{s}(\mathbf{r})$ coupled to the
normal charge density $n (\mathbf{r}) = \rho
(\mathbf{r},\mathbf{r})$ and a second field $\Delta
(\mathbf{r},\mathbf{r}')$ coupled to $\sigma
(\mathbf{r},\mathbf{r}')$. These fields contain
exchange-correlation potentials $v_{xc}(\mathbf{r})$ and $\Delta
_{xc} (\mathbf{r},\mathbf{r}')$, respectively, obtained by
requiring the charge density $\rho (\mathbf{r})$ and the anomalous
density matrix $\sigma (\mathbf{r},\mathbf{r}')$ of the
interacting and noninteracting systems to be identical. TDDFT
requires the same conditions to be satisfied for an externally
driven system at all times\cite{Gross}.  The zero-order
approximation in the standard DFT  obtained by neglecting the
exchange-correlation fields results in the Bogoliubov-de Gennes
equations \cite{Grossbook,deGennes} which take into account the
BCS pairing and classical Coulomb correlations at the Hartree
level.

The DFT equations for superconductors are usually formulated for
the Kohn-Sham orbitals in Hilbert space, and constitute a system
of four self-consistent equations for Bogoliubov transformation
coefficients $u(\mathbf{r})$ and $v(\mathbf{r})$, the normal
density $n (\mathbf{r})$, and
 the anomalous density matrix $\sigma(\mathbf{r},\mathbf{r}')$\cite{Grossbook}. An extension of these equations to include
  a magnetic field was proposed as well\cite{Kohn}.

In this article, we present an alternative derivation of the OGK
equations for a Fermi liquid with spontaneously broken symmetry
(FLSBS) based on a generalized coherent state (GCS) ansatz for the
many electron density matrix. Coherent states were first used to
describe anharmonic dynamical systems such as many body
interacting fermions/bosons\cite{review} while preserving some of
the useful properties of the original Glauber's coherent states
for the harmonic oscillator\cite{glauber1,glauber2}. They
encompass the Glauber coherent states as well as the squeezed
states\cite{glauber1,glauber2}. The time-dependent
Hartree-Fock-Bogoliubov (TDHFB) equations were derived for Boson
systems  using the GCS ansatz \cite{Chernyak,ChernyakChoi}. GCS
are particularly suitable for variational dynamics by virtue the
underlying Lie group algebra\cite{review,perelomov}. Using this
ansatz, we derive equations of motion for expectation values of
the normal and anomalous density matrices. We use the
time-dependent variational principle which allows the description
of the many-body system in terms of a small number of parameters
and is formally closely related to classical Poisson bracket
mechanics, i.e. to the variational equations of motion derived
from the minimum action principle. A GCS representation for the
BCS wave function has been used to analyze the coexistence of
superconductivity and ferroelectricity\cite{Birman,Birman2}. We
obtain self-consistent equations of motion for the normal and
anomalous density matrices\cite{Abrikosov} and  derive the
spectrum of collective excitations, the density of the superfluid
component at finite temperatures, and the temperature of the
transition to the superfluid state for homogeneous and
inhomogeneous superconductors.

Reduced descriptions of many-body systems are naturally recast
using density matrices\cite{Mukamel,Coleman}, and we found it
useful to adopt the Liouville space density matrix representation
\cite{Mukamel,Berman_PRA} of
TDDFT\cite{Grossbook,Gross,Chernyak_TDDFT} for the normal and
anomalous density matrices. This requires solving only two
equations of motion for the normal and anomalous density matrices
$\rho(\mathbf{r},\mathbf{r}')$ and
$\sigma(\mathbf{r},\mathbf{r}')$ coupled to two artificial
external fields. These fields contain an exchange-correlation
contribution and guarantee the charge density and the anomalous
density matrix to be exact at all times\cite{Grossbook,Gross}. The
ground state is the stationary solution of these equations of
motion. This is in contrast to the four self-consistent equations
for Bogoliubov coefficients $u(\mathbf{r})$ and $v(\mathbf{r})$
and the charge density  $n(\mathbf{r}) =
\rho(\mathbf{r},\mathbf{r})$ and anomalous density matrix
$\sigma(\mathbf{r},\mathbf{r}')$\cite{Grossbook2,deGennes}. The
density matrix is a two-point function compare to two one-point
functions $u$ and $v$\cite{Grossbook,Gross}, nevertheless, the
computational cost can be reduced. The reason is that the density
matrices have non-vanishing elements only when $|\mathbf{r} -
\mathbf{r}'|$ is less than a coherence size, which is typically
very short. This allows to neglect many density matrix elements,
making its size scaling linear rather then quadratic. The
Liouville space representation provides a clear picture of the
underlying coherence, since it is not possible to include the
coherence size explicitly in the traditional Hilbert space
computations.

Using our GCS ansatz we further define expressions for first-order
adiabatic (time-independent) exchange-correlation contribution to
the order parameter, which include electron exchange in the
spectrum. This provides corrections to the charge density and the
anomalous density matrix of Ref.\cite{Grossbook2} at each order in
the perturbative series for the exchange correlation potential.
The present approach is applicable for superconductors in general,
 but it is particularly relevant to strongly-correlated high-$T_{c}$
superconductors, where the  BCS and Eliashberg theories do not
apply.

\section{The coherent state free-energy}
\label{groundsec}

We start with the many-electron Hamiltonian $\hat {H}$ where the
electron-electron interaction consists of   both Coulomb repulsion
$V(\mathbf{r}-\mathbf{r}') = e^{2}/|\mathbf{r}-\mathbf{r}'|$, and
pairing attraction $W(\mathbf{r}-\mathbf{r}')$ between two
electrons with opposite spins\cite{Grossbook}
\begin{eqnarray}
\label{ham} \hat{H} &=& \sum_{\nu} \int d{\bf r}\,
\hat{\psi}_{\nu}^{\dagger}({\bf r})
   \left( -\frac{1}{2m}\nabla_{\mathbf{r}}^{2} - \mu\right) \hat{\psi}_{\nu}({\bf r})
    \nonumber \\
   &+& \frac{1}{2}\sum_{\nu \nu'}\int d{\bf r}\,\int d{\bf r}'\,\hat{\psi}_{\nu}^{\dagger}({\bf
r}) \hat{\psi}_{\nu'}^{\dagger}({\bf r}')V(\mathbf{r}-\mathbf{r}')
\hat{\psi}_{\nu'}({\bf r}') \hat{\psi}_{\nu}({\bf r}) \nonumber \\
&-& \sum_{\nu, \nu' \neq \nu} \int d{\bf r}\,\int d{\bf
r}'\,\hat{\psi}_{\nu}^{\dagger}({\bf r})
\hat{\psi}_{\nu'}^{\dagger}({\bf
r}')W(\mathbf{r}-\mathbf{r}')\hat{\psi}_{\nu'}({\bf r}')
\hat{\psi}_{\nu}({\bf r}).
\end{eqnarray}
Here $\hat{\psi}^{\dagger}(\mathbf{r})$ and
$\hat{\psi}(\mathbf{r})$ are the Fermi creation and annihilation
field operators with the anti-commutation relations
$[\hat{\psi}({\bf r}),\hat{\psi}^{\dagger}({\bf r}')]_{+} =
\delta(\mathbf{r} - \mathbf{r}')$ and $[\hat{\psi}({\bf
r}),\hat{\psi}({\bf r}')]_{+} = [\hat{\psi}^{\dagger}({\bf
r}),\hat{\psi}^{\dagger}({\bf r}')]_{+} = 0$, the indices
 $\nu$ and $\nu'$ denote the spin projections; $m$ is the
effective band mass of electron, $\mu$ is the chemical potential
(Fermi energy), and $e$ is the electron charge.

We further expand the field operators in a single electron basis
set $\phi_{i\nu}(\mathbf{r})$
\begin{eqnarray}\label{expbs1}
 \hat{\psi}_{\nu}^\dagger(\mathbf{r},t) = \sum_{i}
\phi_{i\nu}^{*}(\mathbf{r})\hat{a}_{i\nu}^{\dagger}(t); \nonumber
\end{eqnarray}
\begin{eqnarray}\label{expbs2}
 \hat{\psi}_{\nu}(\mathbf{r},t) = \sum_{i}
\phi_{i\nu}(\mathbf{r})\hat{a}_{i\nu}(t) ,
\end{eqnarray}
where $\hat{a}_{i\nu}^{\dagger}$ and $\hat{a}_{i\nu}$ are the
corresponding Fermi operators with the anti-commutation relations
$[\hat{a}_{i},\hat{a}_{j}]_{+} = \delta_{ij}$ and $[\hat{a}_{i},
\hat{a}_{j}]_{+} = [\hat{a}_{i}^{\dagger},
\hat{a}_{j}^{\dagger}]_{+} = 0$; $\phi_{i\nu} ({\bf r})$ are
orthonormal atomic basis functions; and $i$ runs over all basis
electronic orbitals.

Substituting  Eq.~(\ref{expbs2}) in Eq.~(\ref{ham}) gives
\begin{equation}\label{eq-hamilt}
\hat{H} = \sum_{i,j,\nu} t_{ij}\, \hat{a}^\dagger_{i\nu}
\hat{a}_{j\nu} + \sum_{i,j,k,l \atop \nu,\nu'} V_{ijkl}
   \hat{a}^\dagger_{i\nu} \hat{a}^\dagger_{j\nu'} \hat{a}_{k\nu'} \hat{a}_{l\nu}
  - \sum_{i,j,k,l \atop \nu,\nu'\neq \nu} W_{ijkl}
   \hat{a}^\dagger_{i\nu} \hat{a}^\dagger_{j\nu'} \hat{a}_{k\nu'} \hat{a}_{l\nu}.
\end{equation}
 Here $t_{ij}$ is the single-electron matrix element
\begin{equation}\label{eq-1electr}
t_{ij} = \int d{\bf r}\, \phi_i^* ({\bf r})
   \left( -\frac{\hbar^2 \nabla_{\mathbf{r}}^{2}}{2m_b} - \mu\right)
   \phi_j ({\bf r}).
\end{equation}
$V_{ijkl}$ is the Coulomb electron-electron repulsion
\begin{equation}\label{eq-2electrV}
V_{ijkl} = \int d{\bf r}_1 d{\bf r}_2\, \phi_i^* ({\bf r}_1)
\phi_j^* ({\bf r}_2) \frac{e^{2}}{| {\bf r}_1 - {\bf r}_2 |}
   \phi_k ({\bf r}_1) \phi_l ({\bf r}_2) ,
\end{equation}
 $W_{ijkl}$ is an attraction
responsible for the creation of electron Cooper pair.

Eq.~(\ref{eq-hamilt}) describes the interacting many-fermion
system with Coulomb repulsion, and attraction between two
electrons with opposite projections of spin. In ordinary ($BCS$)
superconductors this attraction originates from electron-phonon
interaction\cite{Schrieffer}; in $YBCO$ superconductors the
short-range attraction results from the thermodynamically
equilibrated phase ordering producing charge stripe
order\cite{Carlson}, and it assumes in the following form
\begin{equation}\label{eq-2electrW}
W_{ijkl} = A \int d{\bf r}_1 d{\bf r}_2\, \phi_i^* ({\bf r}_1)
\phi_{j}^* ({\bf r}_2) \left(\frac{|\mathbf{r}_1 -
\mathbf{r}_2|}{r_{0}} \right)^{-n}
   \phi_k ({\bf r}_1) \phi_{l} ({\bf r}_2),
\end{equation}
where $n > 1$ is a positive rational number; $A$ and $r_{0}$ are
constants, determined by  system geometry.

 Our derivation is based on the
following  ansatz for the time-dependent many-electron density
matrix. At zero temperature the system is in  a pure state, and
the density matrix is given by $K(t) \propto |\psi(t)\rangle
\langle \psi(t)|$, where the (unnormalized) many-electron
wavefunction is assumed to be of the form
\begin{eqnarray}
|\psi (t)\rangle &=& \exp \left[ \int dt \int d\mathbf{r}\int
d\mathbf{r}'  \sum_{\nu \nu'} h (\mathbf{r}, \mathbf{r}', t)
\hat{\psi}_{\nu}(\mathbf{r})\hat{\psi}_{\nu'}^{\dagger}(\mathbf{r}')
+ \sum_{\nu, \nu' \neq \nu} \Delta (\mathbf{r}, \mathbf{r}', t)
\hat{\psi}_{\nu}^{\dagger}(\mathbf{r})\hat{\psi}_{\nu'}^{\dagger}(\mathbf{r}')
\right]
|\Omega _{0}\rangle \nonumber \\
&=& \exp \left[\sum_{ij}(\sum_{\nu \nu'} h
_{ij}(t)\hat{a}_{i\nu}^{\dagger}\hat{a}_{j\nu'} + \sum_{\nu, \nu'
\neq \nu} \Delta
_{ij}(t)\hat{a}_{i\nu}^{\dagger}\hat{a}_{j\nu'}^{\dagger}) \right]
|\Omega _{0}\rangle , \label{ansatz}
\end{eqnarray}
with $|\Omega _{0}\rangle$ being an arbitrary single Slater
determinant\cite{Thouless}. Eq.~(\ref{ansatz}) is a generalization
of the RPA and BCS wavefunctions: setting $\Delta_{ij} = 0$ it
reduces to the Thouless representation of the single Slater
determinant\cite{Thouless}, for $h_{ij} = 0$ it reduces to the BCS
ansatz for the superconductor\cite{Schrieffer}.

At finite temperature $T$ our ansatz reads
\begin{eqnarray}
&& K =
\frac{\exp(-\hat{H}_{0}/(k_{B}T))}{Tr\exp(-\hat{H}_{0}/(k_{B}T))}
\, \label{ansatzrho}
\end{eqnarray}
where $k_{B}$ is a Boltzmann constant, and
\begin{eqnarray}
&& \hat{H}_{0} = \int dt \int d\mathbf{r}\int d\mathbf{r}' \left[
\sum_{\nu \nu'} h (\mathbf{r}, \mathbf{r}', t)
\hat{\psi}_{\nu}(\mathbf{r})\hat{\psi}_{\nu'}^{\dagger}(\mathbf{r}')
+ h^{*} (\mathbf{r}, \mathbf{r}', t)
\hat{\psi}_{\nu}^{\dagger}(\mathbf{r})\hat{\psi}_{\nu'}(\mathbf{r}')
 \right. \nonumber \\
&+&   \left. \sum_{\nu, \nu' \neq \nu} \Delta (\mathbf{r},
\mathbf{r}', t)
\hat{\psi}_{\nu}^{\dagger}(\mathbf{r})\hat{\psi}_{\nu'}^{\dagger}(\mathbf{r}')
+ \sum_{\nu, \nu' \neq \nu} \Delta^{*} (\mathbf{r}, \mathbf{r}',
t) \hat{\psi}_{\nu}(\mathbf{r})\hat{\psi}_{\nu'}(\mathbf{r}')
\right] \, \label{ansatzrho1}
\end{eqnarray}
or using our basis set
\begin{eqnarray}
&& \hat{H}_{0} = \sum_{ij}\left[ \sum_{\nu \nu'}
h_{ij}(t)\hat{a}_{i\nu}^{\dagger}\hat{a}_{j\nu'} + \sum_{\nu \nu'}
h_{ij}^{*}(t)\hat{a}_{i\nu}\hat{a}_{j\nu'}^{\dagger} + \sum_{\nu,
\nu' \neq \nu}
\Delta_{ij}(t)\hat{a}_{i\nu}^{\dagger}\hat{a}_{j\nu'}^{\dagger}
\right. \nonumber \\ &+& \left.
 \sum_{\nu, \nu' \neq \nu} \Delta_{ij}^{*}(t)\hat{a}_{i\nu}\hat{a}_{j\nu'} \right]
 \label{ansatzrho2}
\end{eqnarray}
Here $Tr$  denotes the trace in the many electron Fock space. A
similar ansatz was recently used for Bose
condensation\cite{ChernyakChoi}. Eq.~(\ref{ansatzrho}) constitutes
a generalized coherent state (GCS)\cite{Blaizot_Ripka,Bender} (see
Eq.~(\ref{ansatz1})). Eq.~(\ref{ansatz}) is a limiting case of
Eq.~(\ref{ansatzrho}) obtained by a specific choice of
parameters\cite{ChernyakChoi}; Eq.~(\ref{ansatzrho}) thus holds at
finite temperatures as well as at $T = 0$.

The parameters $h_{ij}$ and $\Delta_{ij}$ will be determined
variationally by minimizing the grand canonical free energy
\begin{eqnarray}
\label{freeen} F(\mu, T) \equiv Tr(\hat{H} K) - k_{B}T
Tr(K\log(K)) \equiv \mathcal{H}  - T \mathcal{S} ,
\end{eqnarray}
Here $\mathcal{H}$ is the enthalpy, $\mathcal{S}$ is the entropy,
and the chemical potential $\mu$ controls the average number of
electrons $N$ through the following
constraint\cite{ChernyakChoi,Mermin}
\begin{eqnarray}
\label{contn} Tr\left(K \sum_{i\nu}\hat{a}^\dagger_{i\nu}
\hat{a}_{i\nu}\right) = N .
\end{eqnarray}

Instead of using $h$ and $\Delta$ as the variational parameters,
we shall switch to the following variables: the normal density
matrix
\begin{eqnarray}
\label{ndm} \rho_{i\nu j\nu} \equiv Tr(K \hat{a}^\dagger_{i\nu}
\hat{a}_{j\nu}),
\end{eqnarray}
and the anomalous density matrices
\begin{eqnarray}
\label{andm} \sigma_{i\nu j-\nu} \equiv Tr(K \hat{a}_{i\nu}
\hat{a}_{j-\nu}),
\end{eqnarray}
\begin{eqnarray}
\label{andm1} \sigma_{i\nu j-\nu}^{*} \equiv Tr(K
\hat{a}^\dagger_{i\nu} \hat{a}^\dagger_{j-\nu}).,
\end{eqnarray}

The next step will be to express the free energy in terms of
$\rho$ and $\sigma$. We start with the enthalpy.
 Since $K$ is the exponent of a quadratic operator, we can use
Wick's theorem\cite{Abrikosov} and express all averages of
products of creation and annihilation operators with respect to
$K$ (denoted with a subscript $0$) as averages of pairs of
operators, which are generators of the closed algebra. In
particular we have:
\begin{eqnarray}
\langle \hat{a}_{i}^{\dagger }\hat{a}_{j}^{\dagger
}\hat{a}_{k}\hat{a} _{m} \rangle_{0} &=& - \langle
\hat{a}_{i}^{\dagger }\hat{a}_{k} \rangle_{0} \langle \hat{a}
_{j}^{\dagger }\hat{a}_{m} \rangle_{0}  + \langle
\hat{a}_{j}^{\dagger }\hat{a}_{k} \rangle_{0} \langle
\hat{a}_{i}^{\dagger }\hat{a}_{m}
\rangle_{0} + \langle \hat{%
a} _{i}^{\dagger }\hat{a}_{j}^{\dagger } \rangle_{0} \langle
\hat{a}_{k} \hat{a}_{m} \rangle_{0} . \label{Wick}
\end{eqnarray}
Using this factorization we obtain for the enthalpy (see Appendix
\ref{ap.1pr}), where for brevity we omit spin indices $\nu$:
\begin{eqnarray}
\label{classh} \mathcal{H} \equiv Tr(\hat{H} K) =
\frac{1}{2}\sum_{i,j}\left[\tilde{h}_{ij} (\rho_{ij} -
\rho_{ji}^{*}) + \tilde{\Delta}_{ij}\sigma_{ij}^{*} +
\tilde{\Delta}_{ij}^{*}\sigma_{ji}\right] .
\end{eqnarray}
Here $\tilde{h}$ is a matrix with elements
\begin{eqnarray}
\label{hdef} \tilde{h}_{ij} = t_{ij} +  \frac{1}{2U} \sum_{k,l
\atop \nu,\nu'} (V_{iklj} - \delta_{\nu\nu'}V_{ilkj})\rho_{k\nu
l\nu}.
\end{eqnarray}
$U$ is the volume
 and $\tilde{\Delta}$ is the order parameter matrix  with elements
\begin{eqnarray}
\label{delta1} \tilde{\Delta}_{ij} \equiv  \sum_{mn}
W_{ijmn}\sigma_{mn} .
\end{eqnarray}

We next require the expectation of the effective Hamiltonian
$\hat{H}_{0}$ (Eq.~(\ref{ansatzrho2})) to be the same as the
expectation of $\hat{H}$ (Eq.~(\ref{eq-hamilt})).
\begin{eqnarray}
\label{classh111}  Tr(K\hat{H}) = Tr(K\hat{H}_{0}) .
\end{eqnarray}

It can be easily verified that the condition Eq.~(\ref{classh111})
is met provided we set in Eq.~(\ref{ansatzrho2}) $h_{ij} =
\tilde{h}_{ij}$ (Eq.~(\ref{hdef})) and
 $\Delta_{ij} = \tilde{\Delta}_{ij}$
 (Eq.~(\ref{delta1}))\cite{Blaizot_Ripka}. We next turn to
 computing the enthalpy.
The effective quadratic Hamiltonian $\hat{H}_{0}$ given by
Eq.~(\ref{ansatzrho2}) can be alternatively recast in the form
\begin{eqnarray}
\label{quadH} \hat{H} = tr(\hat{Q}\hat{R}),
\end{eqnarray}
where the matrix $\hat{Q}$ is defined as
\begin{eqnarray}
\hat{Q} \equiv \left (
\begin{array}{cc}
\hat{h} &  \hat{\Delta} \\
- \hat{\Delta}^{*} & - \hat{h}^{*}
\end{array}
\right )  .   \label{QR}
\end{eqnarray}
$\hat{R}$ is the generalized single-particle density matrix
\begin{equation}
\hat{R} \equiv \left (
\begin{array}{cc}
\hat{\rho} & \hat{\sigma} \\
- \hat{\sigma}^* & - \hat{\rho}^* + \openone
\end{array}
\right ) ,   \label{gR1}
\end{equation}
and the symbol $tr$ in Eq.~(\ref{quadH}) stands for the trace in
the single electron space.

 Using the ansatz Eq.~(\ref{ansatzrho}), the
many-electron system described by the Hamiltonian
Eq.~(\ref{eq-hamilt})  can be mapped onto the ideal  system of
non-interacting fermions with the quadratic Hamiltonian
$\hat{H}_{0}$, determined by the matrix
$\hat{Q}$\cite{Blaizot_Ripka} with the parameters defined by
Eqs.~(\ref{hdef}) and~(\ref{delta1}). The entropy $\mathcal{S}$ of
the system of fermions described by the quadratic Hamiltonian
Eq.~(\ref{quadH}) is given by\cite{Blaizot_Ripka}
\begin{eqnarray}
\label{entropy2} \mathcal{S}(\hat{\rho},\hat{\sigma},\mu,T) =
-k_{B} tr[\hat{f}\log\hat{f} + (\openone - \hat{f})\log(\openone -
\hat{f})] ,
\end{eqnarray}
where the matrix $\hat{f}$ is
\begin{equation}
\hat{f} \equiv  \frac{1}{\exp(\hat{Q}/(k_{B}T)) + \openone}  .
\label{eqmf}
\end{equation}

We are looking for  the normal and anomalous density matrices
$\hat{\rho}$ and $\hat{\sigma}$ that minimize the free energy
(Eq.~(\ref{freeen})) together with Eqs.~(\ref{classh})
and~(\ref{entropy2}) assuming the GCS ansatz for the many-electron
density matrix Eq.~(\ref{ansatzrho}). This minimization yields the
following equation\cite{Blaizot_Ripka}
\begin{equation}
\hat{R} =  \frac{1}{\exp(\hat{Q}/(k_{B}T)) + \openone}  ,
\label{eqmR}
\end{equation}
which gives
\begin{eqnarray}
\label{nf} \hat{\rho}  = 1 -
\hat{\varepsilon}[\hat{E}]^{-1}\tanh\left(\hat{E}/(2k_{B}T
)\right),
\end{eqnarray}
\begin{eqnarray}
\label{delta2} \hat{\sigma} =
\frac{1}{2}\hat{\Delta}[\hat{E}]^{-1}\tanh\left( \hat{E}/(2k_{B}T
)\right) ,
\end{eqnarray}
where the the order parameter matrix $\hat{\Delta}$ is defined by
its matrix elements $\Delta_{ij}$ (Eq.~(\ref{delta1})).

 Eqs.~(\ref{nf}),~(\ref{delta2}) and
~(\ref{delta1}) constitute self-consistent equations for the
equilibrium $\hat{\rho}$ , $\hat{\sigma}$ and $\hat{\Delta}$. The
$\hat{E}$ matrix is defined as
\begin{eqnarray}
\label{E1} \hat{E} = \sqrt{\hat{\Delta}^{2} +
\hat{\varepsilon}^{2}},
\end{eqnarray}
where  matrix elements of $\hat{\varepsilon}$ in the single
electron basis set are
\begin{eqnarray}
\label{eps} \varepsilon_{ij} \equiv t_{ij} +
\sum_{kl}\left[\tilde{V}_{iljk} - (1/2)\tilde{V}_{ilkj}
\right]\rho_{kl} ,
\end{eqnarray}
and
\begin{eqnarray}
\label{Vt} \tilde{V}_{iklm} \equiv \frac{1}{2} [V_{iklm} +
V_{kiml}] .
\end{eqnarray}

Substituting Eqs.~(\ref{nf}),~(\ref{delta2}) and ~(\ref{delta1})
into Eq.~(\ref{classh}) gives
\begin{eqnarray} \label{ground}
\mathcal{H}(\hat{\rho},\hat{\sigma},\mu,T) =  tr[2 \hat{E}\hat{n}
+ \hat{\varepsilon} - \hat{E} + \hat{\Delta}\hat{\sigma} -
\frac{1}{2}\hat{M}\hat{\rho}] ,
\end{eqnarray}
where the matrix elements of $\hat{M}$
\begin{eqnarray}
\label{Mij} M_{ij} = \sum_{kl} \left[\tilde{V}_{iklj} -
\frac{1}{2} \tilde{V}_{ikjl} \right]\rho_{kl} ;
\end{eqnarray}
 and the $\hat{n}$ matrix is
\begin{eqnarray}
\label{nf1} \hat{n} \equiv \left[ \exp\left(\hat{E}/2k_{B}T
\right) + \openone \right]^{-1}.
\end{eqnarray}
 At zero temperature  Eq.~(\ref{ground}) gives the ground state energy.

\section{Equations of Motion for Generalized Coherent States}

We are interested in the dynamics of the system coupled to two
external fields: $v_{s}(\mathbf{r})$, which is coupled to the
normal density $\rho (\mathbf{r}) = \rho (\mathbf{r},\mathbf{r})$,
and $\Delta_{ext} (\mathbf{r},\mathbf{r}')$, which is coupled to
the anomalous density matrix $\sigma
(\mathbf{r},\mathbf{r}')$\cite{Oliveira,Gross}. These fields
account for exchange-correlation and provide a starting point for
the TDDFT framework\cite{Gross}. The total Hamiltonian then
becomes $\hat{H}_{T} = \hat{H} + \hat{H}_{ext}$, where $\hat{H}$
is given by Eq.~(\ref{eq-hamilt}), and $\hat{H}_{ext}$ represents
the interaction with the fields\cite{Oliveira,Gross}
\begin{eqnarray}
\label{ham_art} \hat{H}_{ext} &=& \sum_{\nu} \int d{\bf r}\,
v_{ext}(\mathbf{r},t) \hat{\psi}_{\nu}^{\dagger}({\bf r})
   \hat{\psi}_{\nu}({\bf r}) - \sum_{\nu \nu', \nu \neq \nu'}\int d{\bf r}\,\int d{\bf r}'\, \Delta_{ext}(\mathbf{r},\mathbf{r}',t) \hat{\psi}_{\nu}^{\dagger}({\bf
r}) \hat{\psi}_{\nu'}^{\dagger}({\bf r}') \nonumber \\
&-& \sum_{\nu \nu', \nu \neq \nu'}\int d{\bf r}\,\int d{\bf r}'\,
\Delta_{ext}^{*}(\mathbf{r},\mathbf{r}',t) \hat{\psi}_{\nu}({\bf
r}) \hat{\psi}_{\nu'}({\bf r}') .
\end{eqnarray}

Our goal is to compute the dynamics of the system given by the
total Hamiltonian $\hat{H}_{T}$ with the time-dependent external
fields using the GCS ansatz for the many-electron density matrix
Eq.~(\ref{ansatzrho}). This can be accomplished by applying the
closed equations of motion for the averages of GCS generators (see
Eq.~(\ref{variationaleq}) in Appendix \ref{ap.1pr}). These
equations were obtained from  the finite temperature
time-dependent variational
principle\cite{Mermin,perelomov,ChernyakChoi}.
 Since $\rho_{i\nu j\nu}$, $\sigma_{i\nu
j-\nu}^{*}$ and $\sigma_{i\nu j-\nu}^{*}$ are averages of GCS
generators, we can immediately derive closed variational equations
of motion for these quantities in real
space\cite{perelomov,ChernyakChoi}. Substituting the parameters of
the energy $\mathcal{H}$ (Eq.~(\ref{classh})) and $\hat{H}_{ext}$
(Eq.~(\ref{ham_art})) in the closed equations of motion for the
averages of GCS generators Eq.~(\ref{variationaleq}), we get
($\hbar = 1$)
\begin{eqnarray}
\label{rho} i\frac{\partial \rho
(\mathbf{r},\mathbf{r}',t)}{\partial t} &=&
-\frac{1}{2m_{b}}(\nabla_{\mathbf{r}}^{2} -
\nabla_{\mathbf{r}'}^{2})\rho (\mathbf{r},\mathbf{r}',t)
 \nonumber \\
&+& \int d\mathbf{r}_{2}[V(\mathbf{r} - \mathbf{r}_{2}) -
V(\mathbf{r}'
- \mathbf{r}_{2})]  \nonumber \\
&& [\rho (\mathbf{r},\mathbf{r}',t)\rho
(\mathbf{r}_{2},\mathbf{r}_{2},t) - \rho
(\mathbf{r},\mathbf{r}_{2},t)\rho (\mathbf{r}_{2},\mathbf{r}',t)] \nonumber \\
&-&  \int d\mathbf{r}_{2}[W(\mathbf{r} - \mathbf{r}_{2}) -
W(\mathbf{r}' - \mathbf{r}_{2})] \nonumber \\
&& [\sigma^{*} (\mathbf{r},\mathbf{r}_{2},t)\sigma
(\mathbf{r}_{2},\mathbf{r}',t) - \sigma
(\mathbf{r},\mathbf{r}_{2},t)\sigma^{*}
(\mathbf{r}_{2},\mathbf{r}',t)] \nonumber \\
&+& [v_{ext}(\mathbf{r},t) - v_{ext}(\mathbf{r}',t)]\rho
(\mathbf{r},\mathbf{r}',t);
\end{eqnarray}
\begin{eqnarray}
\label{sigma*} i\frac{\partial \sigma^{*}
(\mathbf{r},\mathbf{r}',t)}{\partial t} &=&
-\frac{1}{2m_{b}}(\nabla_{\mathbf{r}}^{2} -
\nabla_{\mathbf{r}'}^{2})\sigma^{*}(\mathbf{r},\mathbf{r}',t)
 \nonumber \\
&+& \int d\mathbf{r}_{2}[V(\mathbf{r} - \mathbf{r}_{2}) -
V(\mathbf{r}'
- \mathbf{r}_{2})]  \nonumber \\
&& [\sigma^{*} (\mathbf{r},\mathbf{r}',t)\rho
(\mathbf{r}_{2},\mathbf{r}_{2},t) - \sigma^{*}
(\mathbf{r},\mathbf{r}_{2},t)\rho(\mathbf{r}_{2},\mathbf{r}',t)] \nonumber \\
&-&  \int d\mathbf{r}_{2}[W(\mathbf{r} - \mathbf{r}_{2}) -
W(\mathbf{r}' - \mathbf{r}_{2})] \nonumber \\
&& [\sigma^{*} (\mathbf{r},\mathbf{r}_{2},t)\rho
(\mathbf{r}_{2},\mathbf{r}',t) - \rho
(\mathbf{r},\mathbf{r}_{2},t)\sigma^{*}
(\mathbf{r}_{2},\mathbf{r}',t)]  \nonumber \\
&+& [\Delta_{ext}(\mathbf{r},\mathbf{r}',t) -
\Delta_{ext}(\mathbf{r},\mathbf{r}',t)]\sigma^{*}
(\mathbf{r},\mathbf{r}',t).
\end{eqnarray}

The order parameter $\Delta(\mathbf{r},\mathbf{r}',t)$, which
characterizes the excitation gap in the spectrum (Eq.~(\ref{E1})),
 is defined as
\begin{eqnarray}
\label{delta} \Delta(\mathbf{r},\mathbf{r}',t) \equiv  \int
d\mathbf{r}_{2}W(\mathbf{r} - \mathbf{r}_{2})\sigma
(\mathbf{r}_{2},\mathbf{r}',t).
\end{eqnarray}
It is interesting to note that the stationary solution of
Eqs.~(\ref{rho}) and~(\ref{sigma*}) gives the normal
 $\rho(\mathbf{r},\mathbf{r}')$ and the anomalous density matrix
$\sigma(\mathbf{r},\mathbf{r}')$, which minimize the equilibrium
 free energy Eq.~(\ref{freeen}) in the ground state (Eqs.~(\ref{nf}) and~(\ref{delta2}))\cite{ChernyakChoi}.
The calculation of the free energy in Sec. \ref{groundsec} is thus
not necessary for the present derivation. The formalism of
Appendix \ref{ap.1pr}) allows us to proceed directly from the
ansatz Eq.~(\ref{ansatzrho}) to Eqs.~(\ref{rho})
and~(\ref{sigma*}). The calculations of Sec. \ref{groundsec}
provide a consistency check and connect our results with the more
conventional derivations.

DFT is usually formulated in Hilbert space and involves the
solution of four self-consistent equations for the Bogoliubov
transformation coefficients $u(\mathbf{r})$ and $v(\mathbf{r})$,
and the charge density $n(\mathbf{r}) =
\rho(\mathbf{r},\mathbf{r})$, and the anomalous density matrix
$\sigma(\mathbf{r},\mathbf{r}')$ (Eqs.~(4.121),~(4.129)
and~(4.130) in \cite{Grossbook}). In the presence of the external
time-dependent field, the ordinary TDDFT framework involves also
the solution of a system of four self-consistent equations for
Bogoliubov transformation coefficients $u(\mathbf{r},t)$ and
$v(\mathbf{r},t)$, the local anomalous density $\sigma
(\mathbf{r},t)$ and the current density $\mathbf{j}(\mathbf{r},t)$
(Eqs.~(20),~(21) and~(22) in \cite{Gross}). Here, in contrast, we
obtain the ground state free energy and density matrices
$\rho(\mathbf{r},\mathbf{r}',t)$ and
$\sigma(\mathbf{r},\mathbf{r}',t)$ by the stationary solution of
two equations Eqs.~(\ref{rho}) and~(\ref{sigma*}). The solution of
these equations gives the charge density $n(\mathbf{r},t) \equiv
\rho (\mathbf{r},\mathbf{r},t)$  and the anomalous density matrix
$\sigma (\mathbf{r},\mathbf{r}',t)$.

Eqs.~(\ref{rho}) and~(\ref{sigma*}) unify several widely used
equations: if we neglect the pairing attraction $W = 0$ and set
$\sigma (\mathbf{r},\mathbf{r}',t) =
\sigma^{*}(\mathbf{r},\mathbf{r}',t) =
\Delta(\mathbf{r},\mathbf{r}',t) = 0$ the last integral in r.h.s.
of Eq.~(\ref{rho}) vanishes, and Eq.~(\ref{rho}) reduces to the
standard RPA equation \cite{Thouless}. By neglecting the Coulomb
repulsion $V = 0$, we obtain the BCS equations \cite{Schrieffer}
(where in the r.h.s. of Eqs.~(\ref{rho}) and~(\ref{sigma*}) the
first integrals vanish). Neglecting the second term  in the r.h.s.
of Eq.~(\ref{sigma*}) gives Eliashberg's equations
\cite{Eliashberg}, which incorporate into account Coulomb
repulsion between electrons at the mean field level.

Since our ansatzs for the many-electron density matrix
Eq.~(\ref{ansatzrho}) and wavefunction  Eq.~(\ref{ansatz}) have
the same number of variational parameters $h$ and $\Delta$, they
yield the same equations of motion for the averages of the GCS
generators (i.e. normal and anomalous single-electron density
matrices). The only dependence on temperature and chemical
potential is through the initial conditions (Eqs.~(\ref{nf})
and~(\ref{delta2})). Eqs.~(\ref{rho}) and~(\ref{sigma*}) conserve
the temperature and chemical potential at all times.

In analogy with the RPA analysis\cite{Thouless}, we can look for a
solution for the density matrices in the form of the following
equations for matrices
\begin{eqnarray}
\label{tddmel} \hat{\rho} &=& \alpha \hat{X}\exp(-i\hat{\omega}t)
 + \alpha^{*} \hat{Y}^{*}\exp(i\hat{\omega}^{*} t); \nonumber \\
\hat{\sigma} &=& \alpha \hat{\tilde{X}}\exp(-i\hat{\omega} t) +
\alpha^{*} \hat{\tilde{Y}}^{*}\exp(i\hat{\omega}^{*} t).
\end{eqnarray}
Substituting Eq.~(\ref{tddmel}) into Eqs.~(\ref{rho})
 and~(\ref{sigma*}),
we obtain the  spectrum of the collective excitations
\begin{eqnarray}
\label{col_sp}
 \hat{\omega} = \sqrt{\hat{\Delta}^{2} +
\hat{\varepsilon}^{2}} ,
\end{eqnarray}
where $\varepsilon_{kj}$ and $\Delta_{kj}$ are given by
Eqs.~(\ref{eps}) and~(\ref{delta2}), respectively.

\section{Application to The  Kosterlitz-Thouless  Phase Transition}

Recent studies of high temperature
superconductors\cite{Dagotto,Ford,Carlson} show a competition
between two types of interaction between electrons: a pairing
attraction, which makes the spectrum satisfy the Landau criterium
of superfluidity by creating a gap in the excitation
spectrum\cite{Schrieffer},  and Coulomb repulsion which tends to
eliminate the gap thereby destroying the superconductivity. This
competition leads to ``stripe'' high temperature superconductivity
(HTS) in cuprates (e.g., $YBa_{2}Cu_{3}O_{7-\delta}$  ($YBCO$))
above the low oxygen concentration threshold ($\delta \sim 0.20$).

 When $\Delta(\mathbf{r},\mathbf{r}',t) \neq 0$ the
excitation spectrum satisfies the Landau criterium of
superfluidity, and a three-dimensional ($3D$) system becomes
superconducting. However, cuprates ($YBCO$) are two-dimensional
($2D$) structures\cite{Carlson,Varlamov}. The Kosterlitz-Thouless
transition temperature\cite{Kosterlitz,Nelson} to the superfluid
state in a two-dimensional superconductive system is given by
$T_{c} = (\pi \hbar^{2}
n_{s}(T_{c}))/(4k_{B}m_{b})$,\cite{Kosterlitz,Nelson} where
$n_{s}(T_{c})$ is the temperature dependent superfluid density of
the superconductive system, and $k_{B}$ is Boltzmann constant. For
temperatures  close to the phase transition in the mean field
approximation ($\Delta_{00}(T_{c}^{0}) = 0$), which satisfy $T -
T_{c}^{0} \ll T_{c}^{0}$, the superfluid density is $n_{s}(T) = (2
(T_{c}^{0} - T)n_{2D})/(T_{c}^{0})$,\cite{Abrikosov1} where
$n_{2D} = p_{F}^{2}/(2\pi\hbar^{2})$ is the total two-dimensional
($2D$) density of electrons ($p_{F}$ is a Fermi radius). To find
$T_{c}^{0}$ from the condition $\Delta_{00}(T_{c}^{0}) = 0$ one
needs to solve the self-consistent equations
Eqs.~(\ref{E1})-~(\ref{delta2}) to obtain the temperature
dependence of the order parameter $\Delta(T)$. The temperature of
the phase transition can be estimated in the mean field
approximation\cite{Abrikosov1}. For the gap spectrum of collective
excitations Eq.~(\ref{col_sp}) the mean field transition
temperature $T_{c}^{0}$ is $\Delta_{00} = 1.76 k_{B}
T_{c}^{0}$,\cite{Abrikosov1} where $\Delta_{00}$ is the order
parameter at zero temperature. Combining these expressions, we
obtain for  the temperature of the Kosterlitz-Thouless transition,
below which the superconductivity exists,
\begin{eqnarray}
\label{tkt1} T_{c} = \left( \frac{2k_{B}m_{b}}{\pi\hbar^{2}n_{2D}}
+ \frac{1.76 k_{B}}{\Delta_{00}} \right)^{-1}.
\end{eqnarray}
$T_{c}$ can be calculated using  the order parameter obtained by
solving Eq.~(\ref{delta2}).

Since both $T_{c}$ (Eq.~(\ref{tkt1})) and the order parameter
$\Delta$ (Eq.~(\ref{delta2})) decrease with the Coulomb
electron-electron repulsion $V$ (Eq.~(\ref{eq-2electrV})) and
increase with the electron-electron attraction $W$,
Eq.~(\ref{eq-2electrW}), Eqs.~(\ref{tkt1}) and~(\ref{delta2})
allow to study the interplay of Coulomb repulsion and pairing
attraction between electrons in FLSBS\cite{Carlson}. Cuprates have
 two-dimensional ($2D$) structure\cite{Carlson,Varlamov}, where the
electron-electron Coulomb correlations may not be
neglected\cite{BermanLozovik, BermanLozovik1}. The screened $2D$
long-range Coulomb potential \cite{Ando} in the momentum space is
$V_{2D}(p) = (2 \pi e^{2})/(p + \kappa_{2D})$, where $2D$
Thomas-Fermi screening radius is density-independent $\kappa_{2D}
= \hbar^{2}/(2m_{b}e^{2})$. While the $3D$ short-range Coulomb
potential in the momentum space $V_{3D}(p) = (4 \pi e^{2})/(p^{2}
+ \kappa_{3D}^{2})$, where $3D$ Thomas-Fermi screening radius
\cite{Kittel} increases with density $\kappa_{3D} \sim
n_{3D}^{1/3}$ and almost eliminates the Coulomb potential at large
distances ($r
> \kappa_{3D}^{-1}$). Therefore, the Coulomb electron-electron
correlations in Eq.~(\ref{delta2}) are much more important in the
$2D$ order parameter compared to $3D$. The present theory thus
describes the contribution of these correlations to the spectrum
of collective excitations, the order parameter, superfluid
density, the temperature of the Kosterlitz-Thouless transition and
the DFT exchange-correlation potential.

\section{Discussion}

Using the GCS ansatz (Eq.~(\ref{ansatzrho})) for the many-electron
density matrix we have calculated the ground state free energy,
the equations of motion for the normal and anomalous density
matrices and  the quasiparticle  spectrum of superconductors. This
results in a quadratic expression for the energy
Eq.~(\ref{ground}) in the operators $\hat{a}_{i}$ and
$\hat{a}_{i}^{\dagger}$, without using the Bogoliubov algebra of
$u-v$ transformations\cite{Schrieffer}.

The advantage of the Liouville space representation of TDDFT for
the normal and anomalous density matrices is that it only requires
to solve two equations of motions for the normal and anomalous
density matrices Eqs.~(\ref{rho})-~(\ref{sigma*}) coupled to  two
artificial external fields $v_{ext}(\mathbf{r})$ and $\Delta
_{ext}(\mathbf{r},\mathbf{r}')$, which contain
exchange-correlation\cite{Grossbook}, instead of the four
self-consistent equations for Bogoliubov coefficients $u$ and $v$
and density matrices $\rho$ and $\sigma$. The normal
 $\rho(\mathbf{r},\mathbf{r}')$ and the anomalous $\sigma(\mathbf{r},\mathbf{r}')$ density matrices, which minimize the equilibrium
 free energy Eq.~(\ref{freeen})\cite{ChernyakChoi} in the ground state are
simply given by the stationary solution of the
Eqs.(\ref{rho})-~(\ref{sigma*}). And in order to get this ground
state we actually don't need to derive the parameters of the
effective quadratic Hamiltonian Eq.~(\ref{ansatzrho1}) for the
many-electron density matrix, as we did in Sec.\ref{groundsec}.

 Our equations for the total energy, the spectrum of collective
excitations and the gap are written in a general basis set and
therefore apply to both homogeneous and non-homogeneous systems.
In a homogeneous system we can use the plain waves basis, i.e. for
the two-dimensional system the eigenfunctions of a momentum
$\mathbf{p}$ $\phi_{\mathbf{p}}(\mathbf{r}) =
U^{-1/2}\exp(-i\mathbf{p}\mathbf{r})$; where $U$ is the volume.
Since we used a general basis set (not necessarily plane waves)
our results, which contain non uniform normal $\rho
(\mathbf{r},\mathbf{r}')$ and anomalous $\sigma
(\mathbf{r},\mathbf{r}')$ density matrices, should be able to
describe the short-coherence-length superconductors.  For example,
in $YBCO$, where the $1 nm$ coherence length is comparable to the
lattice constant\cite{Carlson}.

Finally, we comment on the connection of our results to the ground
state energies and the collective spectrum of excitations in
superconductors calculated using Density Functional Theory (DFT)
and Time-Dependent Density Functional Theory
(TDDFT)\cite{Oliveira,Kohn,Grossbook,Gross}. Let us consider the
following exchange-correlation potentials that depend on the
density matrix (rather than merely on the charge density):
\begin{eqnarray} \label{tdhfexcorr}
v_{xc}[\rho](\mathbf{r}) &=& - \frac{1}{2}\frac{\delta}{\delta
n(\mathbf{r})}\left. \left[ \int d\mathbf{r}\int d\mathbf{r}'
\frac{\rho (\mathbf{r},\mathbf{r}')\rho
(\mathbf{r}',\mathbf{r})}{|\mathbf{r} - \mathbf{r}'|}\right]
 \right|_{n(\mathbf{r}') = \bar \rho
(\mathbf{r}',\mathbf{r}')} .
\end{eqnarray}
Substitution Eq.~(\ref{tdhfexcorr}) in the DFT equations for the
charge density and the anomalous density matrix
\cite{Oliveira,Grossbook} gives equations Eqs.~(\ref{nf}) and
(\ref{delta2}) for the normal and anomalous density matrices. The
GCS ansatz (Eq.~(\ref{ansatz})) is thus equivalent to TDDFT
provided we use the approximate exchange-correlation potential,
Eq.~(\ref{tdhfexcorr}). To improve this functional, the adiabatic
(time-independent) exchange-correlation potentials can be obtained
using the functional derivatives\cite{Grossbook}
\begin{eqnarray} \label{excorr}
v_{xc}([\rho, \sigma];\mathbf{r}) = \frac{\delta F_{xc}[\rho,
\sigma]}{\delta \rho(\mathbf{r})} ; \hspace{0.5in}
\Delta_{xc}([\rho, \sigma];\mathbf{r},\mathbf{r}') = -
\frac{\delta F_{xc}[\rho, \sigma]}{\delta \sigma^{*}(\mathbf{r},
\mathbf{r}')} ,
\end{eqnarray}
where the exchange-correlation free energy $F_{xc}$ can be
obtained using Feynmann diagrammatic perturbation theory for the
self-energy of the Green function\cite{Abrikosov,Grossbook2}. The
zero-order normal and anomalous density matrices derived using the
present ansatz (Eq.~(\ref{ansatz})) are given by Eqs.~(\ref{rho})
and~(\ref{sigma*}), respectively. The spectrum of the
corresponding Green function is $E_{ij}$ given by Eq.~(\ref{E1})
together with Eq.~(\ref{eps}).

The first-order exchange-correlation contribution to the order
parameter is identical to that of
 Ref.\cite{Grossbook2} for homogeneous superconductors provided we set $\varepsilon_{ij} = t_{ij}$. The
second term appearing in our expression for the energy
$\varepsilon_{ij}$ Eq.~(\ref{eps})
$$
\sum_{kl}\left[\tilde{V}_{iljk} - \frac{1}{2}\tilde{V}_{ilkj}
\right]\rho_{kl}
$$
comes from  the first term $\alpha
_{ij}(t)\hat{a}_{i\nu}^{\dagger}\hat{a}_{j\nu'}$ in the
exponential in our ansatz Eq.~(\ref{ansatz}), which represents
electron-electron exchange. It corrects each order in the
perturbative series of
 Ref.\cite{Grossbook2} for the exchange correlation potential.
This term is absent in the OKG equations which use as a reference
the Bogoliubov-deGennes approximation, which takes into account
the BCS pairing and Coulomb correlations at the Hartree
level\cite{deGennes} and neglects exchange-correlation
potentials\cite{Oliveira,Gross}.

\section*{Acknowledgements}
The support of The National Science Foundation grant
No.~CHE-0132571 is gratefully acknowledged. We wish to thank Dr.
Vladimir Chernyak for most useful discussions.

\appendix


\section{Generalized Coherent State Representation of a Hamiltonian with Electron-Electron Pairing} \label{ap.1pr}

Mathematically, a set of GCS is determined by a Lie group $G$, its
irreducible unitary vector representation $T$ with the space $V$
and a reference state $|\Omega\rangle\in V$. The GCS are then
states that have a form $T(g)|\Omega\rangle$ with $g\in G$, where
$g$ is a set of parameters\cite{Chernyak,perelomov,ChernyakChoi}.

Our ansatz for the many-electron density matrix
(Eq.~(\ref{ansatzrho})) can be expressed as
\begin{eqnarray}
K (t) &=& \frac{1}{Z}\exp \left( \sum_{i}\lambda_{i}\hat{T}_{i}
\right) , \label{ansatz1}
\end{eqnarray}
where the set of numbers $\lambda_{i}$ parameterizes the density
matrix. The operator set $\{ \hat{T}_{i} \} = \{ \hat{a}_{i\pm \nu
}^{\dagger }\hat{a}_{j\pm \nu},\hat{a}_{i\nu}^{\dagger
}\hat{a}_{j-\nu}^{\dagger },\hat{a}_{i\nu}\hat{a}_{j-\nu}, \hat{I}
\}$ ($\hat{I}$ is the identity operator) which forms the Lie group
$G$ is characterized by the commutation relations among the
complete set of operators $\hat{T}_{i}$ necessary for describing
the quantum dynamics of the system:
\begin{eqnarray}
[\hat{T}_{i},\hat{T}_{j}]=\sum_{k}C_{ij}^{k}\hat{T}_{k},
\label{commutator}
\end{eqnarray}
where $C_{ij}^{k}$ are known as the structure constants of the set $%
\{\hat{T}_{i}\} $. Writing $\hat{T}^{(-)}_{ij} \equiv
\hat{a}_{i\nu}\hat{a}_{j-\nu}$, $\hat{T}^{(+)}_{ij} \equiv
\hat{a}_{i\nu}^{\dagger }\hat{a}_{j-\nu}^{\dagger }$,
$\hat{T}^{(z)}_{ij} \equiv \hat{a}_{i\pm \nu}^{\dagger }\hat{a}_{j
\pm \nu} + \frac{1}{2} \delta_{ij}\hat{I}$, and, since
$\hat{a}_{i}$
satisfy Fermi anticommutation rules, we have the closed algebra of generators $%
\{\hat{T}_{i}\} $ with respect to the following commutation rules
\begin{eqnarray}\label{comrules}
&& {\mbox [\hat{T}^{(-)}_{ij},\hat{T}^{(-)}_{i'j'}] } =
{\mbox [\hat{T}^{(+)}_{ij},\hat{T}^{(+)}_{i'j'}] } = {\mbox [\hat{T}^{(z)}_{ij},\hat{T}^{(z)}_{i'j'}]} = 0 ; \nonumber\\
&& {\mbox  [\hat{T}^{(-)}_{ij},\hat{T}^{(+)}_{i'j'}] } =  \delta
_{i-ji'-j'}(1-(\hat{T}^{(z)}_{i i} +
\hat{T}^{(z)}_{-i -i})) ; \nonumber\\
&&{\mbox [\hat{T}^{(z)}_{ij}, \hat{T}^{(-)}_{i'j'}] }  =  -
\delta_{ii'}\delta_{j-j'}\hat{T}^{(-)}_{j-j'} -
\delta_{ij'}\delta_{j-i'} \hat{T}^{(-)}_{ji'} ;
\nonumber \\
&& {\mbox [\hat{T}^{(z)}_{ij}, \hat{T}^{(+)}_{i'j'}] } =
\delta_{ji'}\delta_{i-j'}\hat{T}^{(+)}_{ij'} +
\delta_{jj'}\delta_{i-i'}\hat{T}^{(+)}_{ii'} ,
\end{eqnarray}

Eqs.~(\ref{ansatzrho}),~(\ref{ansatz1}),~(\ref{commutator})
and~(\ref{comrules}) show that our ansatz for many-electron
density matrix is generated by  the  operator set $\{ \hat{T}_{i}
\}$, which forms the closed algebra of generators with respect to
their binary commutation. The states described by our ansatz
Eqs.~(\ref{ansatzrho})  thus constitute generalized coherent
states (GCS).

The set of generators $\{ \hat{T}_{i} \}$ corresponds to the
Hamiltonian Eq.~(\ref{ham}). The variational equations at zero
temperature are derived as follows: given a Hamiltonian $\hat{H}$,
and time-dependent wave functions $|\Omega (\tau ) \rangle$, we
minimize the action:
\begin{equation}
S[\Omega (\tau )]=\int d\tau \left[ i\left\langle \Omega (\tau
)|d\Omega (\tau )/d\tau \right\rangle -\langle \Omega (\tau
)|\hat{H}|\Omega (\tau )\rangle \right] .  \label{action}
\end{equation}
By choosing $|\Omega (\tau ) \rangle$ to be a GCS, the resulting
variational equations can be written in the
Hamiltonian form for any set  of coordinates $\Omega_{j}$ which parameterize $%
|\Omega \rangle$:
\begin{equation}
\frac{d \Omega_{j}}{d \tau} = \{ {\cal H}, \Omega_{j} \}
\end{equation}
where $\{ \cdots \}$ denote Poisson brackets and ${\cal H}$ is the
classical Hamiltonian defined by:
\begin{equation}
{\cal H}(\Omega) = \langle \Omega | \hat{H} | \Omega \rangle.
\label{av1}
\end{equation}
The Poisson brackets clearly establishes the link between the
variational equations and the classical dynamics.

 When the
classical Hamiltonian is given by
\begin{equation}
{\cal H} = \sum_{n = 1}^{k} \sum_{i_{1} \cdots i_{n}}
h^{(n)}_{i_{1} \cdots i_{n}} \langle \hat{T}_{i_{1}} \rangle
\cdots \langle \hat{T}_{i_{n}} \rangle, \label{cl_ham}
\end{equation}
the Poisson bracket assumes a very simple form provided the wave
functions $| \Omega
\rangle$ are parameterized by the expectation values $\langle \Omega | \hat{T}%
_{j} | \Omega \rangle$ of the operators $\hat{T}_{j}$ rather than
by the parameters $\Omega_j$. These  expectation values then
constitute a full set of parameters that uniquely specify the
quantum state $|\Omega \rangle $. In particular, if $\hat{T}_{j}$
form a closed algebra Eq.~(\ref{commutator}), the Poisson brackets
for $\langle\hat{T}_{j}\rangle$ is given by:
\begin{equation}
\{ \langle\hat{T}_{m}\rangle, \langle\hat{T}_{n}\rangle \} = i
\sum_{k} C_{m,n}^{k} \langle\hat{T}_{k}\rangle,
\label{commutators}
\end{equation}
and the variational equations of motion for
$\langle\hat{T}_{m}\rangle$ take the closed form
\begin{equation}
i\frac{d \langle \hat{T}_{m} \rangle }{d \tau} = \sum_{n = 1}^{k}
\sum_{j=1}^{n} \sum_{i_{0} \cdots i_{n}} C^{k}_{mi_{0}, \cdots,
i_{j}} h^{(n)}_{i_{1} \cdots i_{n}} \langle \hat{T}_{i_{1}}
\rangle \cdots \langle \hat{T}_{i_{n}} \rangle,
\label{variationaleq}
\end{equation}
Using Eq.~(\ref{ansatz}), it therefore suffices to derive the
equations of motion for the expectation values $\langle
\hat{a}_{i}^{\dagger}\hat{a}_{j} \rangle$ and $\langle
\hat{a}_{i}\hat{a}_{j} \rangle$ to uniquely specify the dynamics
of  electrons. This may be done using the differential property of
the Poisson brackets:
\begin{equation}
\{f, gh \} = - \{ gh, f\} = \{ f, g \}h + g \{ f, h \}.
\label{differential}
\end{equation}
When the expectation values are for generators of the set of GCS
of some Lie group $G$, their Poisson brackets are given by the
commutators of the underlying generators of the group. This direct
correspondence between ordinary quantum mechanical commutators and
the Poisson brackets greatly simplifies the calculation, since the
variational procedure is then equivalent to the Heisenberg
equations of motion. Eqs.~(\ref{rho}) and~(\ref{sigma*}) were
obtained using Eq.~(\ref{variationaleq}).


\end{document}